\documentclass{ws-p8-50x6-00}

\def\gtorder{\mathrel{\raise.3ex\hbox{$>$}\mkern-14mu
             \lower0.6ex\hbox{$\sim$}}}
\def\ltorder{\mathrel{\raise.3ex\hbox{$<$}\mkern-14mu
             \lower0.6ex\hbox{$\sim$}}}

\begin{document}

\title{Are Recoil Polarization Measurements of $G_E^p/G_M^p$
Consistent with Rosenbluth Separation Data?}

\author{J. Arrington}

\address{Argonne National Laboratory, Argonne, IL, USA}


\maketitle

\abstracts{Recent recoil polarization measurements in Hall A at Jefferson Lab
show that the ratio of the electric to magnetic form factors for the proton
decreases significantly with increasing $Q^2$.  This contradicts previous
Rosenbluth measurements which indicate approximate scaling
of the form factors ($\mu_p G_E^p(Q^2) / G_M^p(Q^2) \approx 1$).  The
cross section measurements were reanalyzed to try and understand the source
of this discrepancy.  We find that the various Rosenbluth measurements are
consistent with each other when normalization uncertainties are taken into
account and that the discrepancy cannot simply be the result of errors in one
or two data sets.  If there is a problem in the Rosenbluth data, it must be a
systematic, $\epsilon$-dependent uncertainty affecting several experiments.}

\begin{figure}[thb]
\begin{center}
\epsfxsize=16pc 
\epsfbox{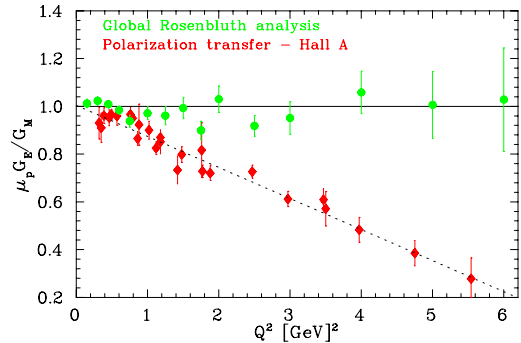} 
\caption{Ratio of electric to magnetic form factor from a global analysis of
cross section data (circles) and from the JLab measurements of recoil
polarization (diamonds).
\label{fig.gegm}}
\end{center}
\end{figure}

The structure of the proton is a matter of universal interest in nuclear and
particle physics.  The electromagnetic structure of the proton can be
parameterized in terms of the electric and magnetic form factors, $G_E(Q^2)$
and $G_M(Q^2)$, which can be measured in elastic electron-proton scattering.  The electric and
magnetic form factors can be separated using the Rosenbluth technique\cite{rosen},
or by measurements of the polarization transfer to the struck nucleon\cite{recoilpol}.
Figure~\ref{fig.gegm} shows the ratio of $\mu_p
G_E/G_M$ as a function of $Q^2$ for the Jefferson Lab recoil polarization
measurements\cite{halla} and from a global Rosenbluth analysis of the cross
section measurements\cite{walker}.  Clearly we must understand this
discrepancy if we want to be confident in our
knowledge of the proton form factors. 

While it is possible that
there is a fundamental problem with one of these techniques, we first want to
understand if we can explain the difference in terms of less fundamental
problems ({\it e.g.} experimental errors or analysis procedures).
The Rosenbluth measurements are
more sensitive to experimental uncertainties as $Q^2$ increases, and
extractions that involve combining multiple data sets are
sensitive to their relative normalization factors.  Thus, we wish to
examine both the individual cross section measurements and the analysis
procedures to see if there could be problems that would explain the
discrepancy between the two techniques.

In the global analysis shown in fig.~\ref{fig.gegm}, many
data sets are combined, and a global fit is performed to extract the
relative normalization of the experiments as well as the value of $G_E$
and $G_M$ at several $Q^2$ values.  Errors in one or more of the experiments or
improper normalization procedures for experiments which combine multiple
measurements could cause such a global fit to give an incorrect result.
In addition, because relative normalization factors are being fit, it
is possible that one could vary the normalization factors for
one or more experiments by an amount that is within the experimental
uncertainty in such a way that the ratio of $G_E/G_M$ changes significantly,
while the overall $\chi^2$ of the fit is not significantly increased ({\it
i.e.} the global minimum might give the results shown in fig.~\ref{fig.gegm},
but a local minimum may give a global fit that is almost as good in which the
ratio of $G_E/G_M$ falls with $Q^2$).

A new global fit was performed in order to investigate possible problems
in the previous data or analyses.  Experiments where multiple spectrometers
were used to take portions of the data were broken up, so that there were
16 data sets (and 16 normalization parameters) for the 13 experiments included.  
As this analysis was focussed on the discrepancy at larger $Q^2$, data
below $Q^2=0.3$ GeV$^2$ were excluded.  The small angle
data ($\theta < 15^\circ$) from the Walker measurement~\cite{walker} were also excluded, because
a later SLAC experiment found corrections that had been neglected in the
analysis\cite{priv}.  The new fit gives results that were similar to
the global analysis by Walker, and no data set had an anomalously large
contribution to the $\chi^2$.
Additional fits were performed with individual data sets left out, to see if the result
might be driven by a single (potentially bad) data set.  No single experiment
had a large impact on the overall fit, and
even removing the three data sets that had the largest effect
only decreased the ratio of $G_E/G_M$ by $\sim$5-10\% at high $Q^2$.

While improvement to the global analysis and removal of data sets did not
allow for agreement between the two techniques, there is still the question
as to whether a different solution for the relative normalizations could
be found which brings the experiments into agreement without significantly
decreasing the quality of the fit.  This was tested in two different ways.
First, $G_M$ was fit to the data, with the ratio of $G_E/G_M$ determined
from a parameterization of the recoil polarization data
($\mu_p G_E/G_M= 1 - 0.13Q^2$).
Even though the normalization parameters for the different data sets are
allowed to vary in this procedure, the overall quality if fit is much lower
when the fit is forced to match the $G_E/G_M$ ratio from the recoil
polarization measurements (the total $\chi^2$ increases by 69
for the fit to 301 cross section data points).
Fixing the ratio of $G_E/G_M$ to match the recoil
polarization measurements gives these data more impact on the fitting than
they should have and ignores their uncertainties, so this test likely overestimates
the inconsistency.  A global analysis including both the cross section data
and the $G_E/G_M$ polarization measurements from fig.~\ref{fig.gegm}
(including their statistical and systematic errors) also gives a significantly
worse overall fit, though not as bad as when the ratio is fixed in the fit
(16 data points are added to the fit and the total $\chi^2$ increases by 49).

Finally, it has been noted\cite{halla} that individual extractions of $G_E/G_M$ from
different cross section measurements are inconsistent.  However, these extractions
often involve combining two or three data sets that cover different
$\epsilon$ ranges, which requires determining the
cross-normalization between experiments.  While various procedures have
been used to determine these normalization factors, the uncertainty
in the normalization is often not taken into account in extracting $G_E$ 
and $G_M$, even though a normalization error can yield a correlated change 
in the ratio at all $Q^2$ values.
Thus, it is difficult to verify the consistency of the underlying cross section data
based on these extractions.  If one examines only experiments where a
single detector covered an adequate range
of $\epsilon$ to perform a Rosenbluth separation, these experiments
are consistent with each other and give results similar to the previous
global fits (although with significantly reduced precision).  One can
increase the amount of data available by including experiments where
multiple detectors were used, but where direct cross-calibrations were
possible within the experiment.  Again, this set of experiments give
consistent results, and are in good agreement with the cross section global
analysis.  The inconsistency of the Rosenbluth extractions appears to come
from the assumptions made when combining data sets at different $\epsilon$
values, and does not indicate a fundamental inconsistency between the
different measurements.

Even if the recoil polarization result is correct and the problem lies with the
cross section data, we must still understand
the problem with the Rosenbluth measurements.  If the recoil polarization data
is correct, this implies that there is a problem in the cross section
measurements that introduces a systematic $\epsilon$-dependence in
multiple data sets.  Even with perfect knowledge of $G_E/G_M$, we need
these cross sections to extract the absolute values of $G_E$ and $G_M$, and
we cannot extract precise and accurate values for the form factors if we
do not know what the problem is with the cross section measurements.

In conclusion, the disagreement between the recoil polarization and Rosenbluth
measurements cannot be explained by assuming that there is a problem with one
or two data sets, nor can they be made to agree by simply adjusting the
relative normalization factors in a global analysis (without significantly
worsening the quality of the fit).  There is no evidence of problems within any
of the data sets (with the exception of the low angle Walker data), and the
existing Rosenbluth measurements are completely consistent. The extractions of
$G_E$ from these data are only inconsistent when one includes analyses that
combine different data sets without properly taking into account the uncertainties in
the relative normalizations.  Thus, there is no experimental evidence to
tell us which of these techniques is failing.  It is important to determine
which is correct not only because we want to know the form factors of the
proton, but also because these techniques are used in other measurements, and
a fundamental problem with either technique could affect other measurements.
Future measurements at JLab including a high precision Rosenbluth\cite{e01001}
separation and a new recoil polarization measurement\cite{hallcrecoil} (using a
different experimental setup) will help us understand the discrepancy and
determine if it is a fundamental problem with one of the techniques or a problem
with the existing data.

This work is supported (in part) by the U.S. DOE, Nuclear Physics Division,
under contract W-31-109-ENG-38.


\begin{thebibliography}{99}


\bibitem{rosen} M. N. Rosenbluth {\it et al.}, Phys. Rev. {\bf 79}, 615 (1950).

\bibitem{recoilpol} R. G. Arnold, C. E. Carlson, and F. Gross, Phys. Rev. {\bf C23}, 363 (1981).

\bibitem{halla} M. K. Jones {\it et al.}, Phys. Rev. Lett, {\bf 84}, 1398 (2000) ;
O. Gayou {\it et al.}, Phys. Rev. {\bf C64} 038292 (2001).
O. Gayou {\it et al.}, Phys. Rev. Lett, {\bf 88} 092301 (2002).

\bibitem{walker} R. C. Walker {\it et al.}, Phys. Rev. {\bf D49}, 5671 (1994).

\bibitem{priv} R. C. Walker, C. E. Keppel, and A. F. Lung, private communications.

\bibitem{e01001} JLab E01-001, J. Arrington and R. E. Segel spokespersons.

\bibitem{hallcrecoil} JLab E01-109, E. J. Brash, C. Perdrisat and V. Punjabi Spokespersons.

\end{thebibliography}
\end{document}